\pdfoutput=1

\documentclass{article}    

\usepackage{authblk}
\usepackage{cite}
\usepackage{graphicx}          
\usepackage{geometry}
\usepackage{xcolor}
\definecolor{customblue}{RGB}{0,81,158}
\usepackage{fancyhdr}
\usepackage{amsmath,amssymb,amsthm}
\usepackage{mathtools} % coloneqq, eqqcolon
\usepackage{bm}
\usepackage{pgfplots}  
\pgfplotsset{compat=newest}
\usepackage{tikz, listofitems}
\usetikzlibrary{calc, shapes, backgrounds, bending, decorations.pathmorphing, positioning, decorations.pathreplacing, arrows.meta}
\usepgflibrary{arrows}
\usepackage{subcaption}
\usepackage{enumitem} 
\usepackage{units}
\usepackage{adjustbox}

\usepackage{hyperref}
\hypersetup{colorlinks=true, linkcolor=black, urlcolor=customblue, citecolor=black, anchorcolor=black}
\begin{document}

\renewcommand\Affilfont{\normalsize}
\title{Data Publishing in Mechanics and Dynamics: Challenges, Guidelines, and Examples from Engineering Design}

\author[1]{Henrik Ebel\thanks{The corresponding author is H.~Ebel. Email address: henrik.ebel@lut.fi}}
\author[3]{Jan van Delden}
\author[3]{Timo L\"{u}ddecke}
\author[4]{Aditya Borse}
\author[4]{Rutwik Gulakala}
\author[4]{Marcus Stoffel}
\author[5]{Manish Yadav}
\author[5]{Merten Stender}
\author[6]{Leon Schindler}
\author[6]{Kristin Miriam de Payrebrune}
\author[7]{Maximilian Raff}
\author[7]{C.\ David Remy}
\author[2]{Benedict R\"{o}der}
\author[8]{Rohit Raj}
\author[8]{Tobias Rentschler}
\author[8]{Alexander Tismer}
\author[8]{Stefan Riedelbauch}
\author[2]{Peter Eberhard}

\affil[1]{Department of Mechanical Engineering, LUT University, Finland}
\affil[2]{Institute of Engineering and Computational Mechanics, University of Stuttgart, Germany}
\affil[3]{Institute of Computer Science, University of G\"{o}ttingen, Germany}
\affil[4]{Institute of General Mechanics (IAM), RWTH Aachen, Germany}
\affil[5]{Cyber-Physical Systems in Mechanical Engineering, Technische Universit\"{a}t Berlin, Germany}
\affil[6]{Institute for Computational Physics in Engineering, RPTU Kaiserslautern-Landau, Germany}
\affil[7]{Institute for Nonlinear Mechanics, University of Stuttgart, Germany}
\affil[8]{Institute of Fluid Mechanics and Hydraulic Machinery, University of Stuttgart, Germany}

\date{}
\maketitle
\vspace{-1em}
\begin{abstract}
Data-based methods have gained increasing importance in engineering, especially but not only driven by successes with deep artificial neural networks. 
Success stories are prevalent, e.g., in areas such as data-driven modeling, control and automation, as well as surrogate modeling for accelerated simulation. 
Beyond engineering, generative and large-language models are increasingly helping with tasks that, previously, were solely associated with creative human processes. 
Thus, it seems timely to seek artificial-intelligence-support for engineering design tasks to automate, help with, or accelerate purpose-built designs of engineering systems, e.g., in mechanics and dynamics, where design so far requires a lot of specialized knowledge. 
However, research-wise, compared to established, predominantly first-principles-based methods, the datasets used for training, validation, and test become an almost inherent part of the overall methodology. %, and the disparate kinds of data employed in different engineering sub-disciplines needed for holistic design are usually non-trivial to understand, visualize, and represent appropriately to work well with machine-learning methods. 
Thus, data publishing becomes just as important in (data-driven) engineering science as appropriate descriptions of conventional methodology in publications in the past. 
This article analyzes the value and challenges of data publishing in mechanics and dynamics, in particular regarding engineering design tasks, showing that the latter raise also challenges and considerations not typical in fields where data-driven methods have been booming originally. 
Possible ways to deal with these challenges are discussed and a set of examples from across different design problems shows how data publishing can be put into practice. 
The analysis, discussions, and examples are based on the research experience made in a priority program of the German research foundation focusing on research on artificially intelligent design assistants in mechanics and dynamics.
\end{abstract}
\vfill

\section{Introduction}\label{sec:Intro}
In engineering research and practice, in recent years, data-oriented thinking and the usage and research of data-based methods have become much more prevalent, especially in modeling, simulation, and control. 
However, engineering research in these fields relies on decades of established processes regarding research communication, publishing, and methodology that have been geared toward classical, first-principles approaches. 
One manifestation of this is that, in areas that were pioneering the latest surge in data-based approaches, i.e., mostly subfields of computer science, quite different customs regarding publications are now established. 
There, datasets and executable code are often openly available with articles. 
Value is put on open review processes and speedy dissemination of research results, where relevance is preferably ensured through convincing performance in benchmark problems and executable code that can be tested by readers. 
Yet, this need not mean that copying these customs to data-based research in engineering is the way to go. 
Engineering problems have their own character, and it may be that some of the traditional customs of disseminating engineering research have value in their own right even in new contexts. 
After all, engineering research nowadays often uses data-based components, but still also relies on classical methodologies, like first-principles modeling and analytical argumentations or even proofs where it is feasible and successful. 

The described way of thinking has also emerged with researchers partaking in the German Research Foundation's priority program "Daring More Intelligence -- Design Assistants in Mechanics and Dynamics" (SPP 2353), which unites at least thirteen senior researchers and a similar number of research associates in the common endeavour to leverage artificial intelligence to assist in designing the mechanics and dynamics of the systems of the future. 
During their collaborative and diverse work on problems from engineering design, it became evident that some common thought needed to be put on data publishing and sharing. 
Most importantly, it appeared that dealing with mechanics and dynamics and, in particular, with engineering design leads to specific challenges regarding data. 
This article is the result of discussions in the priority program and of the experience from nine collaborative research projects therein. It serves three key purposes. 

Firstly, the article shall raise awareness to the challenges of data publishing in mechanics and dynamics and system design, in particular for engineering scientists also embarking on more data-driven research, and it shall present some possibilities and guidelines to deal with these challenges. 

Secondly, the article presents a set of seven examples from the aforementioned research projects that outline how the described challenges of data publication were overcome. 
The authors hope that these concrete examples will further nurture ongoing discussions and change within the community and will be of help to other scientists in mechanics and dynamics who want to publish data and data-driven research.

Thirdly, while not the immediate goal, the article and its examples may show to scientists from outside engineering that very challenging and worthwhile problems exist in mechanics, dynamics, and engineering design, where good data-driven contributions could lead to sustained technological progress. 
The authors invite data scientists to try their methods on the published data. 

In general, data publishing and also scientific treatments of how and why to publish data are emerging topics. 
However, the authors are not aware of comparable treatises specifically for mechanics and dynamics and, therein, the topic of engineering design. 
Still, many existing works and guidelines are also relevant to the field looked at here, and will also partly appear in the following, such as, e.g., the well-known FAIR principles on data publishing~\cite{WilkinsonEtAl16}. 
Interestingly, some proposed and recognized guidelines on how to document datasets in the machine-learning community take inspiration from practices established in engineering industries~\cite{GebruEtAl21}, used there to document technical components. 
Thus, they should feel natural to scientists in engineering. 
Moreover, whereas this article will exemplify realizations of data publishing at the examples of the authors' datasets from the studied engineering problem-fields, there is a growing number of published datasets in closely related fields, e.g., from natural sciences. 
In particular, this holds true for datasets containing large libraries of solutions of (often expensive) numerical simulations, guided by the hope to use them to train ML-based solvers that can be queried to more quickly arrive at approximate solutions of expensive-to-solve differential equations. 
A well-known example is the PDEBENCH~\cite{TakamotoEtAl24}, but also datasets specific to computational fluid dynamics may be found~\cite{ElrefaieEtAl24,YagoubiEtAl24}, which is a field with notoriously costly simulations that are hence particularly worthwhile to expedite or circumvent. 
A rarer dataset that is directly meant for design applications is presented in~\cite{RegenwetterCurryAhmed21}, containing a large number of human-designed bicycle models. 
Similarly,~\cite{NobariEtAl22} describes a dataset consisting of 100 million planar linkage mechanisms, which can be used for data-driven kinematic design. 

At the same time, an increasing number of methodological works exist that deal with generative design in engineering, see, e.g., the review paper~\cite{RegenwetterHeyraniFaez22}, giving an overview of various generative methods used for engineering design. 
Therein, the authors identify several challenges for deep-learning based design methods, including for instance whether deep-learning based approaches can actually be creative enough to produce novel results as well as the usability and feasibility of resulting designs, e.g., for production. 
Unlike the present work,~\cite{RegenwetterHeyraniFaez22} is focused on the learning methods themselves, not on the aspect of data publishing, its challenges, and implications for the proceedings of science on generative engineering design in mechanics and dynamics. 
A  review of machine-learning methods commonly used in computational mechanics, not necessarily for design tasks, is given by \cite{HerrmannKollmannsberger24}. 
In general, the breadth of work and methodology discussed in~\cite{RegenwetterHeyraniFaez22,HerrmannKollmannsberger24} can be seen as a motivation for this work's approach to look at data-driven design in mechanics and dynamics from a data-centric instead of a method-focused or application-centric perspective. In particular, the article focuses on what that entails for publishing practices. 

The article is organized as follows. 
Based on a general motivation in Section~\ref{sec:motivation}, challenges for publishing data in mechanics, dynamics and for associated design tasks are formulated in Section~\ref{sec:challenges}. 
Ideas and guidelines how these challenges might be addressed are discussed in Section~\ref{sec:solutions}. 
Section~\ref{sec:examples} presents examples of published datasets, program code, and model problems, showing first attempts to meet and overcome the aforementioned challenges, and concludes with lessons learned from the examples. 
Finally, Section~\ref{sec:conclusions} draws conclusions and gives and outlook with regard to the findings specific for engineering design. 

\section{The Value of Publishing Data}\label{sec:motivation}
Traditionally, with classical methods of mechanics and dynamics, calls for the publication of data and code were rooted in the desire to enable a straight-forward reproducibility of results in the spirit of good science. 
However, if described well enough, proposed methods could also be tried and checked, sometimes quite readily, by implementing and testing them on problems with known reference solutions. 
It was simple to separate the proposed methodology from input data, such as initial conditions of a simulation.  
In contrast to that, for data-based methods, the lines between methodology and data can blur significantly. Put bluntly, training data can become an integral part of the methodology itself~\cite{RechtEtAl19,SunEtAl17}. 
After all, there can be a certain interdependency between the makeup of the dataset, such as the selection of data and its mathematical representation, and the quality of the results obtained with a certain data-based method. 
In many applications, coming up with a suitable amount of data with the required properties can be much harder than the setup of the learning method and the execution of the training process. 
Therefore, in such scenarios, the principles of good science require publishing datasets together with the proposed methodology or algorithm. For this, it is adequate to publish the data exactly in the way the (also to be published) code demands. 
To people more familiar with 'traditional' first-principles methods, this may seem like a nuisance brought up by data-driven methods. 

However, when the publication of data is thought in a broader fashion, it can present significant opportunities for research. 
If properly published, data can serve as an abstraction layer. This enables researchers not familiar with the concrete problem behind the data to propose data-driven solutions, based on abstract properties of the data and the problem, accelerating scientific progress. 
The underlying problem in question, as represented by the data, can then become a benchmark or model problem used by data scientists to test and develop new methods. 
The term benchmark problem should be understood slightly differently than it has sometimes been used in mechanics and dynamics. 
In classical dynamics, it can mean a certain kind of physical problem that every simulator should be able to solve accurately to be deemed 'correct' in the sense that it can reproduce an accepted ground-truth solution.  
Instead, subsequently, benchmark problems are to be understood as prototypes or archetypes for typical problems from engineering (design), prepared in a way suitable for data-based methods and data scientists. 
In particular, there need not be a ground-truth solution. 
At best, methods performing well with the benchmark problem may be good candidates for the whole class of problems represented by the benchmark.  

It is to be expected that the availability of data and well-prepared benchmark problems can very much influence and steer in which areas progresses in machine learning are made. However, an analysis of published datasets and openly accessible and executable benchmark problems shows that there are comparatively few from the field of mechanics and dynamics, especially when focusing on published material suitable for people that are not experts in the field. 

Obviously, apart from the kind of data mentioned before, there have always been published datasets that are not meant to be interesting primarily due to their value as a basis for machine learning. For instance, measurements of temperatures at a certain geographic location or measurements of certain material properties are among these datasets for which the provision of the data itself is, at least classically, the purpose of the investigation and publication. This kind of data publishing is not the main purpose of this paper.  

\section{Challenges in Data Publishing}\label{sec:challenges}
Some challenges to be faced for the publication of data are quite common beyond the specific characteristics of the research on design assistant systems. These include:
\begin{itemize}
\item \textbf{Interpretability and Domain Knowledge:} Raw data can be tough to understand for anybody unfamiliar with the generation process of the data. 
Furthermore, the representation of the data may heavily depend not only on the underlying system but also on the chosen analysis method and observation approach. 
In simulations, the deformation of a simple mechanical beam, for example, can be computed using numerical discretization schemes that involve the partitioning of the physical domain into small elements. 
The shape and/or numbering of those elements may however greatly vary across different modeling or simulation approaches, while displaying the same or a very similar elastic deflection of the beam. 
Comparability of different simulation techniques -- and the resulting data -- can be difficult for the very same physical system. 
This can happen analogously in experiments, if measurements are taken at different points of the same mechanical object or with different measurement hardware, which may come with different built-in filtering, e.g., for noise reduction. 
In the light of this challenge, it is maybe not surprising that many of the first, renewed successes of machine learning in recent years happened for data interpretable without special education, like text, speech, and image data. 
\item \textbf{Problem Complexity:} There can be conflicting interests regarding a desirable level of the studied problem's complexity, and depending on it, it may be hard to place publications that focus on the proposition of a new model problem and accompanying datasets. 
In mechanics and dynamics, this type of publication is less common than in data science, and the usual conventions in the field may prefer engineering problems at the upper end of the complexity scale, deeming them publishable even outside any machine-learning context. 
However, overly complex scenarios significantly complicate the explanation of scenarios and data, limiting the appeal to machine learning experts and thus the potential data-scientific impact.
\item \textbf{Generalizability:} Compared with language models or computer vision, problems from mechanics and engineering design and corresponding collected data can appear (or sometimes are) overly specific in the eyes of a researcher from, e.g., the machine learning community, even in cases where the domain expert may know of the potentially large impact of a solution of the underlying engineering problem or, more generally, of its generalizability to other problems. Hence, published datasets may remain unused even if they were easily usable and understandable. 
\end{itemize}

\noindent Beyond these, engineering design also presents a very specific challenge, namely the challenge of \textbf{evaluation}. 
In realistic engineering design problems, there is typically not a "correct" reference solution. There usually is no known, globally optimal design for predefined optimality criteria, even if there is an agreed-upon single optimality criterion. 
In contrast, the goal is to obtain a "better" design than initially given or than achieved with other design methods. This is in stark contrast to benchmark problems widely popular in supervised machine learning, where a labeled set of test data can be used to assess the performance of the learning method using simple calculations. 
In engineering design, the evaluation process of a candidate design can be very intricate, involving advanced simulations that may require extensive amounts of computation, potentially on high-performance general-purpose computers or specialized architectures. 
Often more critically, even installing and running the software environment for these test runs may be a burden only domain experts can overcome. Hence, means to judge the quality of results need to be published in an accessible fashion.  

\section{Dealing with the Challenges}\label{sec:solutions}

The following ideas and approaches may help to overcome the identified challenges.
 
\vspace{0.5em}
\noindent\textbf{Evaluation:} If the evaluation mechanism happens to be lightweight and easy to install and use, it can be sufficient to distribute it with the dataset or model problem. 
In other cases, a workable, yet more involved, solution is to publish a (web-)interface to an evaluator, to which results can be uploaded. The evaluation is performed on the server, and the evaluation results are made available to the user. The advantage is that the server and evaluation toolchain are set up and configured only once by the domain experts. 
Luckily, the required technology is already available from other domains, albeit for slightly different reasons. 
In data science, it is sometimes even undesirable to publish test datasets used for evaluation since then researchers may tune their method also with the test dataset. 
For these reasons, there exists open-source software that greatly helps to implement such online-evaluation or benchmarking tools. 
A popular example is \href{https://github.com/codalab/codabench}{Codabench}~\cite{XuEtAl22}, which is even used to do public competitions for defined tasks. Some challenges, like the funding and maintenance of the server, remain. However, a growing number of success stories shows that this can work and can significantly help to popularize a benchmark problem, particularly in the crucial phase when the problem is new. 

To find new, interesting papers, data scientists nowadays often browse through websites such as \href{https://paperswithcode.com/}{Papers with Code}~\cite{MetaAI23}, where papers appear jointly with their open-source code and their performance in public benchmark problems. 
There, papers are not ranked by traditional measures, such as the prestige of the journal they appear in, but based on benchmark results as well as on ratings of users, which may base their rating on their experience when executing the code. State of the art approaches are \href{https://paperswithcode.com/sota}{grouped by tasks} (closely related to the type of data such as time series, images, etc.). 
The visibility of papers is dictated by the performance on known benchmark problems rather than on their number of citations. Thus, papers without code or benchmark results will most likely remain invisible there, highlighting the culture change from the 'traditional' ways of mechanics and dynamics. 
Another popular platform to share datasets and trained machine-learning models is \href{https://huggingface.co/}{Hugging Face}~\cite{HuggingFace23}, where grouping the available datasets by tasks again reveals a focus on natural language processing and computer vision, and  a severe lack of mechanics, dynamics, and engineering design examples. 

\vspace{0.5em}
\noindent\textbf{Interpretability and Domain Knowledge:} Some of the issues involved are appropriately dealt with by established principles for data management, such as the FAIR principles (short for findable, accessible, interoperable, reusable, \cite{WilkinsonEtAl16}), where some of those principles can be fulfilled by including descriptive metadata with the data. 
There exist professionally curated data repositories implementing these principles, some operated by universities (e.g., the \href{https://darus.uni-stuttgart.de/}{DaRUS} repository, see~\cite{UniStuttgart23}) or by third parties. There also exist registries of research data repositories, e.g., \href{https://www.re3data.org/}{re3data}~\cite{KIT23}. 
For engineering and engineering design problems, mere metadata, such as describing what is stored in which format, may still not be enough to facilitate usage, especially for researchers from other domains. 
Hence, at the very least, datasets should come shipped with some open-source code that, as a form of minimal example, reads the data from the data files and visualizes it in an intuitive way. 
The latter is very important in the field of mechanics and dynamics, as a good visualization of the data can be quite insightful – compare, e.g., an explanatory spatial animation of motion to just supplying raw data or line plots in a variety of different coordinate frames. If data and program code are published at different places, hyperlinks should point from one to the other, as well as to and from associated research articles. 
A further way to increase accessibility and reduce the inevitable technical hurdle to look at published code and data is technology such as \href{https://colab.research.google.com/}{Google Colab}~\cite{Google23}, which permits very straightforward (Python) code execution via a web browser, including access to GPUs. 
An explanatory worksheet, which outlines the idea behind the problem, interactively visualizes the data, and explains the basic usage of the proposed methodology, can help greatly to interest and make familiar interested parties with a specialized problem, as the execution of the worksheet is just one hyperlink and one click away. 
Technology, regulations, and guidelines for research data management advance swiftly. Even binding legislation may appear in the future~\cite{DFG23}. Many universities now have strategies for the management of research data that provide useful information. Moreover, helpful resources include the German Research Foundation's \href{https://www.dfg.de/en/principles-dfg-funding/basics-and-principles-of-funding/research-data}{comments on the topic}~\cite{DFG23b}, including possibilities to receive funding for data management, which supplement the usual \href{https://www.dfg.de/en/principles-dfg-funding/basics-and-principles-of-funding/good-scientific-practice}{guidelines for good scientific practice}~\cite{DFG23c}. 

\vspace{0.5em}
\noindent\textbf{Conflicting Interests on Complexity:} Before publishing data or a model problem, one needs to decide on the goal of the data publication, including who is the target audience (closely connected to the venue of publication), and what is the benefit of finding new, better methods to deal with the data. 
The complexity of the problem should then be calibrated so that it is engaging to the target audience in its difficulty, where difficulty may mean different things for different audiences. 
Still, domain experts may not see the value in handling a problem that they deem "simple" (e.g., already solved by "established" methods) with new methods from machine learning. This risk can be alleviated by an appropriate framing of the problem as well as an explicit explanation of the benefits of finding new methods and of the perspectives to move to more intricate problems when the simpler ones have been solved with data-driven approaches. 
Luckily, engineering design processes deliver plenty of natural motivation, e.g., because traditional methods currently quickly lead to computational intractability, especially when combining multiple design aspects. 

\vspace{0.5em}
\noindent\textbf{Generalizability:} To alleviate concerns regarding the specificity of tasks, once again, a proper framing of the problem is important. This is often glossed over in domain-specific publications, as, for instance, the significance of typical fluid-dynamics model problems is clear to mechanical engineers. 
In contrast, to have a data-scientific impact, the motivation behind and the characteristics of engineering tasks needs to be explained in a way engaging to researchers usually dealing, e.g., with language or image processing. Guiding questions may be: Why is the problem of general importance? What makes the problem engaging from a data-centric perspective? 
How are the mechanical and dynamical intricacies of the problem represented in the specific properties of the data? Are there problem-specific requirements, e.g., on the inputs and outputs or on the mapping to learn? 
When purely looking at the data and its properties, are there similarities to data from other tasks and domains? Where are interesting differences?

\section{Examples: Published Code and Data}\label{sec:examples}
Subsequently, published data and code, as they have resulted from the priority program "Daring More Intelligence -- Design Assistants in Mechanics and Dynamics" (SPP 2353) of the German Research Foundation, are listed and discussed in the light of the aforementioned findings, to serve as inspiration how some of the formulated challenges can be overcome, and to learn which types of data and problems appear in mechanics, dynamics, and engineering design.

\subsection{Vibrating Plates}\label{sec:ex_vibrating_plate}
Noise and vibrations of mechanical structures such as machinery and vehicles impact the human well-being and health~\cite{basner2014auditory}, which could be mitigated with the aid of design assistants aimed at reducing noise and vibrations. 
Motivated by this, the vibrating plates dataset addresses the design of plates with indentations to minimize vibrations.
In mechanical structures, vibrations emit noise and can be transmitted from sound sources like engines to, for example, the passenger cabin in a car. 
Indentations in plate-like structures can help reduce vibrations, if they are well-placed.
 
 \begin{figure}[h]
 \centering
 \includegraphics[scale=1.2]{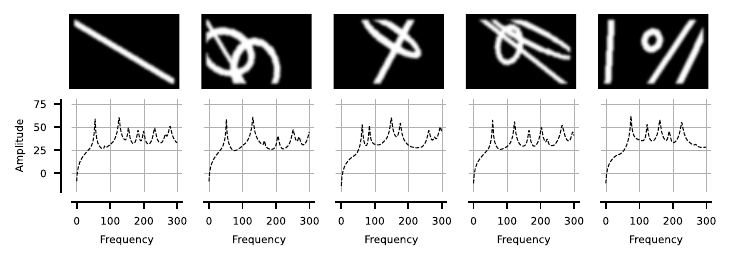}
 \caption{Example plates from the vibrating plates dataset along with the averaged mean squared velocity of the vibrations. Indentations are marked in white in the top row}
 \label{fig:vibrating_plates}
 \end{figure}

Figure~\ref{fig:vibrating_plates} visualizes typical pieces of input and output data. On the input side, the dataset consists of image-type data describing the indentation patterns and scalar parameters describing the material as well as geometric properties of the plate. 
On the output side, the dataset describes the dynamic response of the plate at discrete excitation frequency steps. Here, different fidelity resolutions are provided: Image-type data of the velocity field represent high-fidelity data, whereas the domain averaged mean squared velocity represents an integral quantity. Regarding \textbf{problem complexity}, this dataset exhibits many representative characteristics of a vibroacoustic problem, but the choice of indented plates instead of more complex geometries enables straightforward application of many standard machine learning methods. The \href{https://doi.org/10.25625/UWF7RB}{dataset} is published in a curated data repository from the University of Göttingen~\cite{DeldenEtAl23}, the \href{https://github.com/ecker-lab/Learning\_Vibrating\_Plates}{code} is published in a repository on GitHub~\cite{DeldenEtAl23b}. The data is stored in a common format (hdf5) compressed with a publicly available compression algorithm.
 
The dataset can be used to train machine-learning-based regression models. In particular, the regression model should learn the mapping from beading patterns and plate properties to the velocity field. Regarding  \textbf{evaluation}, metrics are defined in the paper to evaluate the performance of the regression model on a separate test dataset, e.g., the classical mean squared error in the dynamic response or the error in resonance peak location. Results for several deep-learning-based methods are provided.
 
Regarding \textbf{interpretability}, the \href{https://doi.org/10.25625/UWF7RB}{dataset}~\cite{DeldenEtAl23}, the associated paper~\cite{delden2023vibroacoustic}, and the accompanying \href{https://github.com/ecker-lab/Learning\_Vibrating\_Plates}{code}~\cite{DeldenEtAl23b} are designed to be accessible to non-domain experts who have a background in machine learning. To achieve this, the descriptions in the paper avoid domain-specific language where possible and important concepts are explained within it. Also, care was taken to follow conventions for machine learning papers and code repositories. This includes a focus on reproducibility and employing standard convenience tools like conda environment definitions, jupyter notebooks and a clear and detailed ReadMe file in the code repository. In addition, video visualizations of the dataset are displayed directly in the ReadMe and can be produced with a provided notebook. As a verification step, a researcher from a different institute with no prior knowledge was asked to download and visualize the data, and was able to do this within 10 to 15 minutes. Based on this dataset, two additional publications have been accepted, dealing with extending the dataset~\cite{schultzDAGA} and employing guided diffusion to targeted design~\cite{delden2024minimizing}.

\subsection[Crashworthiness of Crash Boxes]{Database of Crashworthiness Analysis of Crash Boxes}
\begin{figure}[h]
    \centering
    \includegraphics[width=0.73\columnwidth,trim={1cm 2cm 2cm 4.6cm},clip]{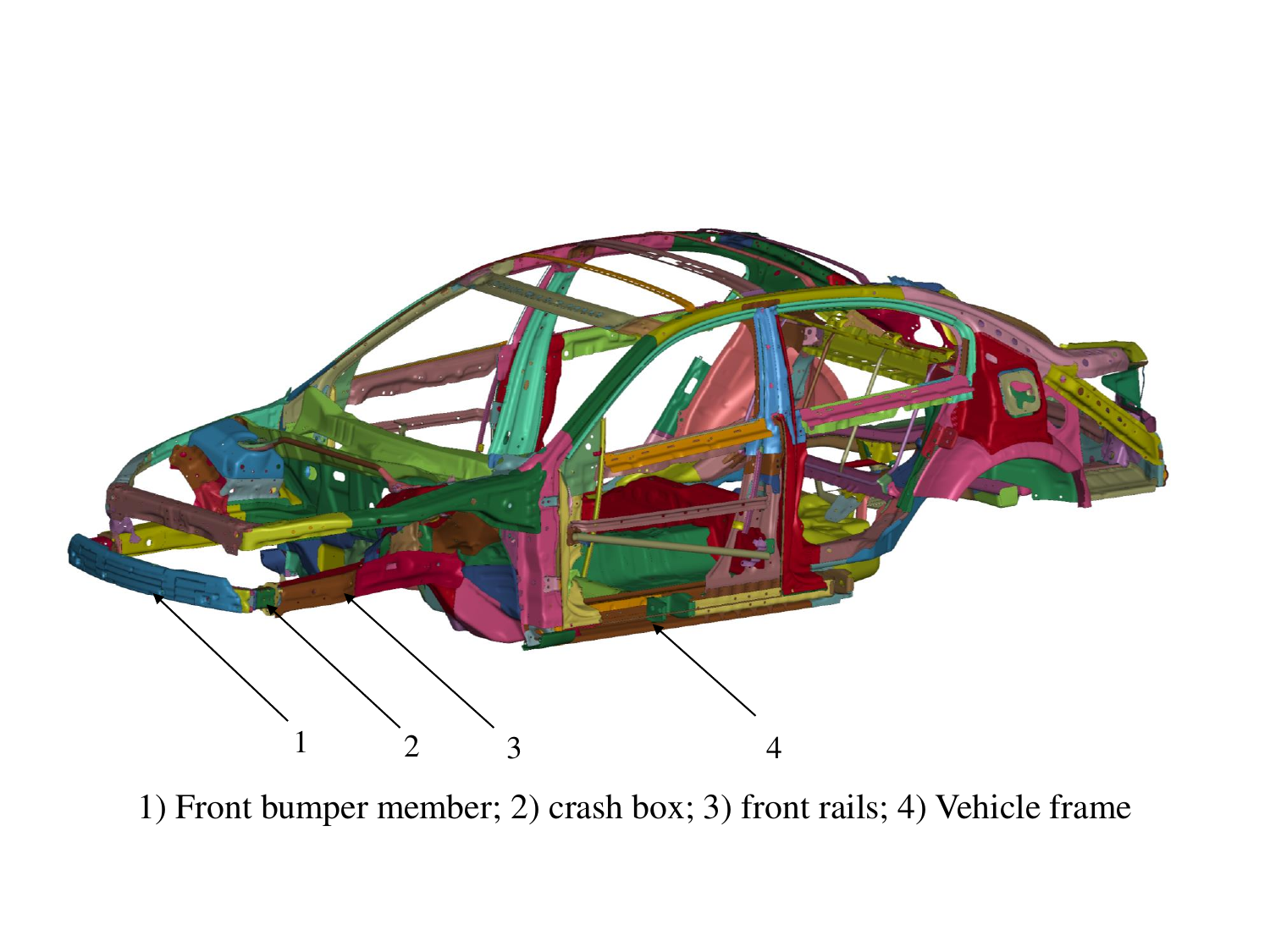}
    \caption{Crash box position in automobiles}
    \label{fig:crashbox_position}
\end{figure}%
\noindent Every year, approximately 2.54 million people in Germany alone are involved in automotive crashes, which makes vehicle crashworthiness a crucial part of the development of automobiles~\cite{destatis}. To help reduce the seriousness of the impact and prevent serious injuries and fatalities, both active and passive crash management systems are installed in automobiles. Passive safety systems are usually structural components which are built to absorb impact energies and reduce potential harm to the occupants. Since these systems are static, they must be meticulously designed and optimized to perform in various scenarios. One such passive safety system is a crash box behind the front bumper, as shown in Figure~\ref{fig:crashbox_position}. 
The above model is derived from the Honda Accord Model provided in the NHTSA database \cite{nhtsa_crash_model}. 
Crash boxes are thin-walled hollow structures that are installed between the subframe and the front bumper. These absorb the kinetic energy from the impact by undergoing axial deformation and local buckling, thereby reducing the forces experienced by the occupants. The optimal design of the crash box is crucial since frontal impacts are the most common impact type in road traffic. 

In this context, a publicly accessible database has been established that contains valuable information on various crash box configurations and their performance. This database is not merely a result of an associated research project; it has been developed to provide other researchers with valuable insights and tools. The database is publicly available in a \href{https://github.com/MrAdityaBorse/FEM_Data_Impact_Simulation/}{GitHub repository}~\cite{Borse24EtAl_Repo}, see also the corresponding results published in the article~\cite{Borse_PAMM23}. The detailed information on the database is presented subsequently. The provided data enables scientists and engineers to:

\begin{itemize}
    \item \textbf{Compare different methods:} Researchers can utilize the data to compare their optimization methods against the results from the analysis from \cite{Borse_PAMM23}, fostering constructive dialogue about different approaches to improve crash safety. Researchers can utilize the data to investigate classical optimization methods such as evolutionary algorithms, response surface methods (RSM) or modern random search algorithms, which is one example of utilizing the data for a comparative study.

    \item \textbf{Verify their own implementations:} The detailed simulation results offer a solid foundation for testing and validating individual models or simulation techniques in crash analysis. 

    \item \textbf{Develop new approaches:} Access to the data allows others to develop innovative methods or adapt existing techniques to better meet specific requirements of their research projects.

    \item \textbf{Encourage data sharing and collaboration:} Even if the collection of examples in the dataset is small, it may well inspire others to share their own datasets and consider how these could be beneficial for the community. 
    
\end{itemize}
The database maps the relation between the structural and crash test parameters to that of various crashworthiness metrics used to evaluate the crash box design. In the provided database, crash box parameters (width, thickness, and length), crash test parameters (impactor mass, impact velocity) and time required for simulations are recorded. Multiple crash box configurations are evaluated by varying the thickness of the crash box and impact velocity. The objective crashworthiness parameters are peak impact force, energy absorbed, maximum deformed length, mean contact force and mass of the crash box. They are calculated for every configuration based on the FE simulation results. The neighboring components define maximum peak impact force as they should not buckle before the crash box, and generally, the energy absorbed should be maximized. 
\begin{figure}[btp]
    \centering
    \includegraphics[width=0.5\columnwidth]{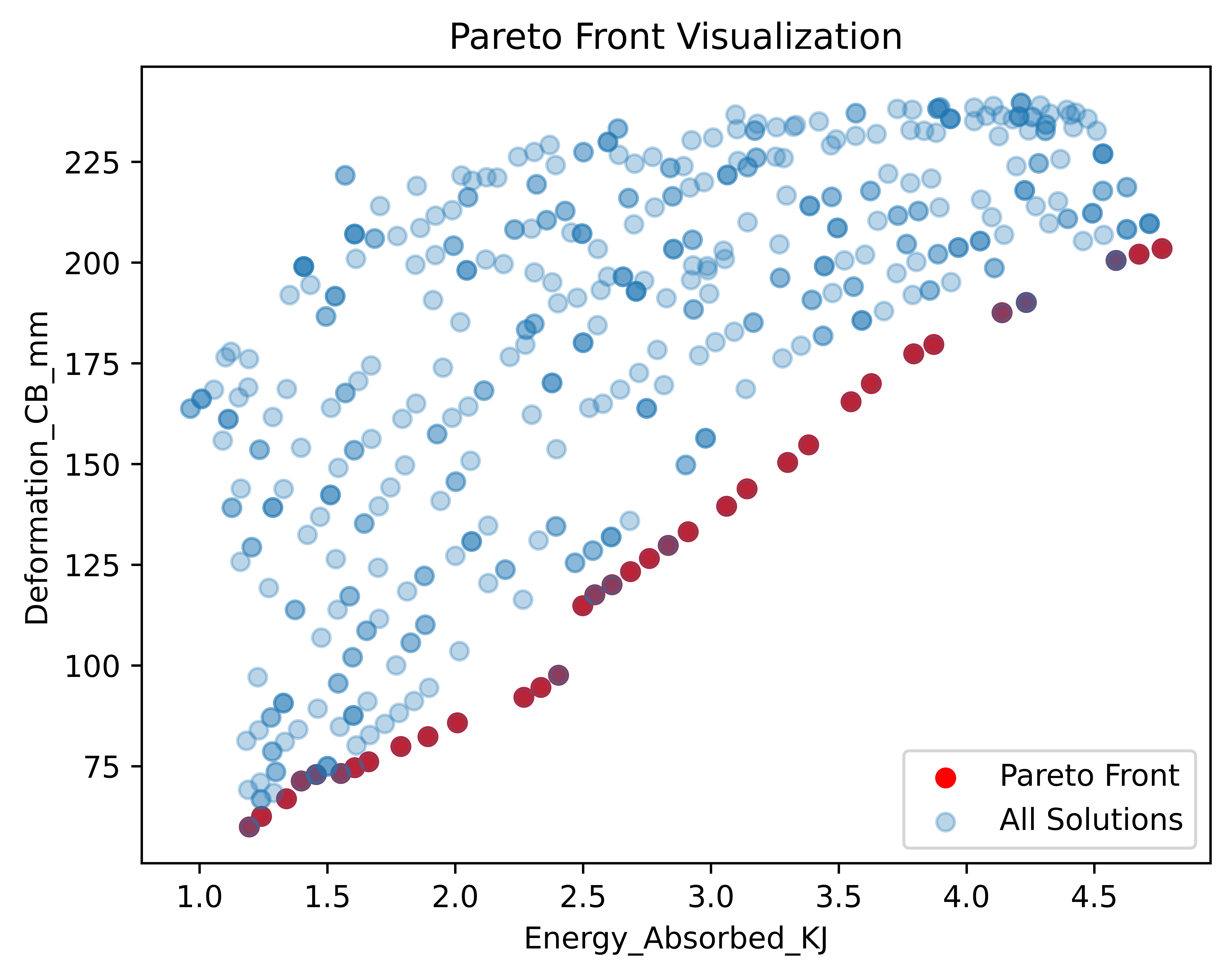}
    \caption{Pareto front for the objectives: maximum deformed length and total energy absorbed}
    \label{fig:pareto_front}
\end{figure}%
Other parameters, such as maximum deformed length, mean contact force, and mass of the crash box, are informative parameters of the crash box's overall performance. Assuming the energy absorbed has to be maximized while reducing the maximum deformed length leads to certain non-dominated solutions, as shown in Figure~\ref{fig:pareto_front}. 
These form the so-called Pareto front, and it denotes the solutions where both objectives are important. 

This database is utilized in a multi-objective optimization of the crash box design to determine the optimal thickness for an ideal peak impact force and allowed maximum deformed length of the crash box~\cite{Borse_AoM24,Borse_PAMM23}. A list of all the dependencies required for executing the scripts is mentioned in the Git repository.

\subsection[Soft Robot]{Backbone Reconstruction of a Non-slender Soft Robot}
\begin{figure}[b]
    \centering
    \includegraphics[width=0.495\linewidth, keepaspectratio]{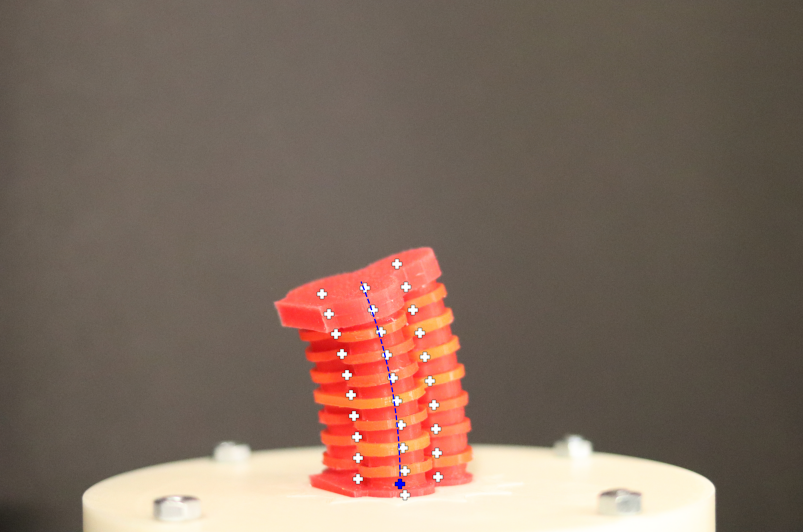}
    \includegraphics[width=0.495\linewidth, keepaspectratio]{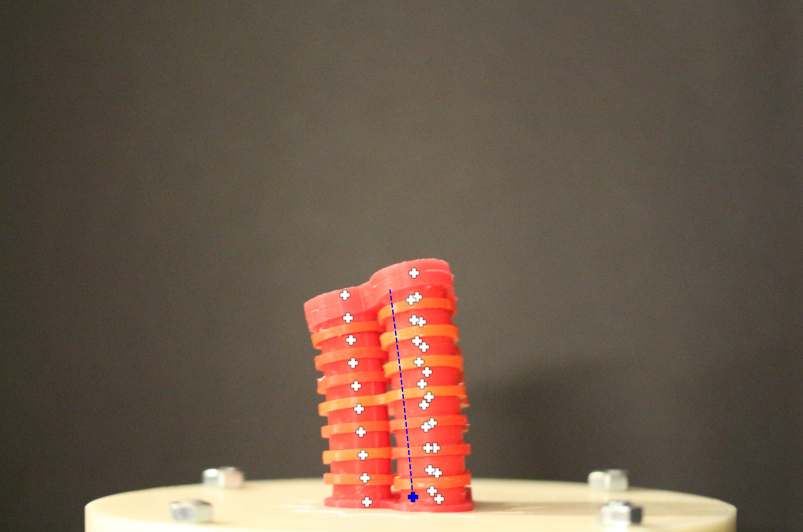}
    \caption{Images of a non-slender soft robot annotated with the reconstructed backbone (blue dashed line) and estimate points (white markers). Taken from~\cite{Schindler2024b} by CC BY}
    \label{fig:reconstructed_backbone}
\end{figure}

The research field of soft robotics is based on the investigation of robots made up of elastic soft materials like silicone.
The use of such elastic soft materials results in a continuous deformation of the robot under load or even under its dead weight.
However, compared to conventional robots made up of rigid connections, the computation of this continuous deformed state can be more involved, but is a crucial part of the control of the soft robot and for the verification of models. 
The so-called backbone can be seen as a one-dimensional description of the continuous deformation of the soft robot, as visualized in Figure \ref{fig:reconstructed_backbone} for a deformed soft robot. 
Especially when manufacturing soft robots by hand, variances in the geometry and the manufacturing accuracy change the behavior significantly. 
Hence, being able to identify the parameters of the soft robot via image processing is a fast and easy way to enhance the reliability of subsequent modeling and to validate simulation results of the soft robot with geometrically accurate beam models. 
To that end, first, a stereo camera system is set up and calibrated. 
The soft robot is then actuated with different pressures while both cameras capture images to obtain the raw data.
The \href{https://zenodo.org/records/11352739}{dataset} contains the calibration data, raw images, and images annotated with the reconstructed backbones. 
The corresponding parameters of the reconstruction are available in~\cite{Schindler2024b}.
Figure~\ref{fig:reconstructed_backbone} displays an annotated entry from this dataset.
The blue, dashed line represents the reconstructed backbone, while the white markers indicate the estimated points used in the optimization process.

Applying the image-based method published by \cite{Hoffmann2024}, with modifications described in \cite{Schindler2024a}, we adapt the technique for reconstructing the backbone of a non-slender soft robot. The original method is provided as the Python software package \textit{icpReconstructor}, and the modifications and extensions for a modular bending actuator, along with a short example, can be found in the associated \href{https://github.com/MKHoffmann/icpReconstructor}{GitHub repository}~\cite{Hoffmann24EtAl_Repo}, as detailed in \cite{Hoffmann2024}. 
The \href{https://zenodo.org/records/11352739}{image dataset} is available in~\cite{Schindler2024b}.

In the dataset, the raw images are annotated to face the issue of \textbf{interpretability} of the results.
Since the optimization for the reconstruction of the backbone operates in pixel space and can be depicted in both figures at each iteration step, the results can be interpreted and even evaluated visually, allowing the practitioner to verify the inner workings of the reconstruction method. 
Furthermore, with a rendering of a deformed geometry using the reconstructed backbone from the dataset, the \textbf{evaluation} can be further supported.
While the reconstruction method can be implemented in software, its usefulness is only leveraged with data captured in experiments. Therefore, the dataset can be seen as part of the modified model.  
Even if the adaptation contained in the dataset is specific to the employed soft robot, it can be viewed as an example of how well the method introduced in~\cite{Hoffmann2024} can be \textbf{generalized} to different tasks. 
The use cases are not limited to soft robots, but can be modified to fit different applications.

A single soft robot bending actuator, as shown in Figure~\ref{fig:reconstructed_backbone}, has limited applicability on its own, as it is usually part of a larger soft robot system. However, focusing on a single module reduces the \textbf{problem complexity} that arises when considering an entire soft robot system with multiple actuation parameters and potential continuous deformations, and makes the reconstruction method more feasible. Nevertheless, the deformation resulting from the pressurization of one or more chambers of the soft robot with or without external loads can be evaluated and compared with the beam deflection.
The described datasets of stereo images \textbf{offer numerous further possibilities for the analysis of soft robots}. Besides the original detection of the backbone, the dataset can be used to analyze and identify parameters of different mechanical properties of the soft robot, e.g., axial stiffness, bending stiffness and torsional stiffness, where the pressure actuation and resulting deformations have to be put in relation. In addition, the behavior of the robot can be evaluated for different material models, providing valuable insights for optimizing design and performance. These applications can significantly improve the understanding and development of efficient and adaptable soft robotic systems.

\subsection[DORA]{Duffing Oscillator Response Analysis (DORA)}
\label{sec:DORA}
In mechanical engineering, the state-of-the-art system design paradigms are mostly based on a number of crucially simplifying and idealizing assumptions on dynamic loads and system complexity. In particular, the non-stationarity of the system dynamics is often regarded as perturbation to some steady-state operation. However, static load cases actually occur only rarely in real-life systems: aircraft engines, wind turbines and vehicle components consistently operate under non-stationary, non-periodic, multi-scale, multi-physical, or in total \textbf{complex} external loads and control inputs. Hence, the transient dynamics of engineering structures must be taken into account early in the design process. At the same time, data acquisition can be expensive, rendering the available sequential \textbf{data sparse and small}: typically, only few operating conditions and corresponding system responses are available. This situation requires models that \textbf{generalize} very well to unseen system parameters or external loads, and which are able to predict complex dynamic behavior. The \textbf{problem complexity} thus arises from four core aspects: (a) the problem involves time-dependent multivariate quantities on input and response sides, (b) many variable system parameters and external loads that affect the system response, (c) critical transitions occur in the system response due to minor changes in the system parameters, and (d) limited observability and small data in the form of time series. \textbf{Interpretability} is not at the core of this modeling task, however highly welcome to elaborate how a model picks up transient behavior.   

\begin{figure}[btp]
    \centering
    \includegraphics[width=0.8\columnwidth]{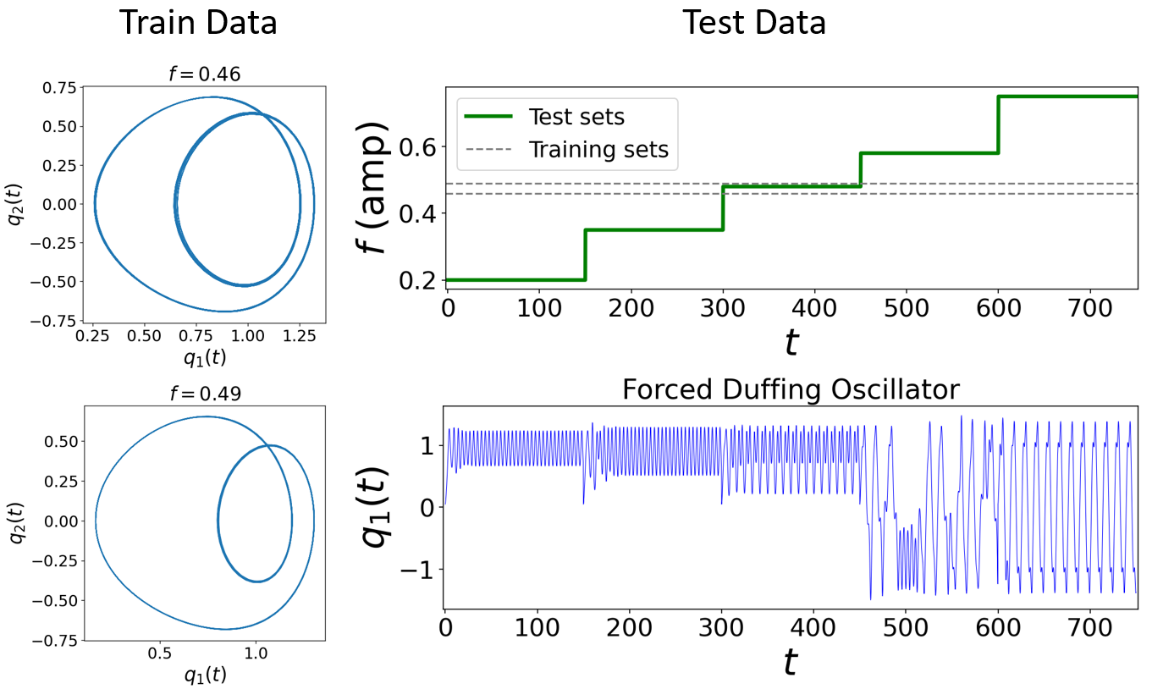}
    \caption{DORA time series prediction task in the small-data limit: the training data (left) comprises two trajectories of period-2 cycle dynamics for two values of external forcing amplitudes. The modeling tasks aims at generalization to different forcing amplitudes that induce qualitatively different dynamics, among others chaotic dynamics, as displayed in the right}
    \label{fig:berlin1}
\end{figure}

The dataset presented in this section deals with the aforementioned challenges in the following ways: it provides data in the form of time series of a minimal mechanical oscillator, namely the Duffing oscillator. This model is complicated enough to show all qualitatively possible deterministic dynamics (regular to irregular) with transitions governed by parametric changes. Moreover, using a synthetic data approach, it is possible to generate training datasets of varying size, thereby allowing to scale the complexity stemming from data availability. The Duffing oscillator represents an externally forced nonlinear oscillator with dynamics
\begin{align}
\dot{q}_1&=q_2,\\
\dot{q}_2&=-c\, q_2 - k\, q_1 -\beta\, q_1^3+f \cos(\omega t)
\end{align}
where the parameters are chosen as $c=0.32$, $k=-1$, $\beta=1$, $\omega=1.5$ to provoke a wide range of differing dynamics under changes to the external forcing. Duffing-type oscillators have been studied and implemented in a wide range of physics and engineering applications such as MEMS resonators~\cite{SabarathinamThamilmaran16}, in energy harvesting devices~\cite{MartensWagnerLitak13}, and to study periodic forcing dependent chaotic behavior of nonlinear systems~\cite{Strogatz24}. Various qualitative changes (bifurcations) of the system dynamics are obtained by varying the forcing amplitude ($f$) while keeping the forcing frequency constant. This variation allows to generate period-1 limit cycle solutions for small forcing amplitude, but also chaotic dynamics for larger forcing amplitudes. Rapid changes in the dynamics occur under minimal changes to the forcing amplitude (period-doubling bifurcation route to chaos) as displayed in Figure~\ref{fig:berlin1}.

The modeling challenge takes the general form of multivariate sequence prediction for parameterized dynamics, and requires auto-regressive models that can advance from a given temporal context into the future. The dataset is comprised of time-series data obtained for specific values of the forcing
amplitude ($f=[0.46, 0.49]$, corresponding to period-2 cycle motions). From this dataset, the model is requested
to predict vibration time series for different forcing values far from the training set, spanning values that are multiples smaller or much larger than the training regime. The modeling task hence asks for time series prediction of qualitatively strongly varying dynamics, and extreme out-of-sample \textbf{generalization} from a minimal amount of training data. Successful models will allow to predict the system's transient dynamics across a wide range of forcings, and accurately predict critical transition points (bifurcations).

For \textbf{evaluation} purposes, the success of the prediction model will depend on the accurate capture of the overall dynamics (the \textit{climate}) as well as the short-term fully-resolved temporal dynamics. The system response characteristics are quantified in terms of (mean) amplitude of the system state $q_1^{2}(t)$. The mean squared error (MSE) can be used as an accuracy quantifier to obtain the deviation of predicted system response characteristics relative to the original ones for each of the external forcing amplitudes, given by
\begin{align}
\varepsilon_{\text{A}} &= \text{MSE}\left(\text{Amp}\left(q_{1,\text{prediction}}^{2}(t)\right), \text{Amp}\left(q_{1,\text{original}}^{2}(t)\right)\right),\label{eq:DORA_MSE1}\\
\varepsilon_{\text{M}} &= \text{MSE}\left(\text{Mean}\left(q_{1,\text{prediction}}^{2}(t)\right), \text{Mean}\left(q_{1,\text{original}}^{2}(t)\right)\right)\label{eq:DORA_MSE2}
\end{align}
where $\varepsilon_{\text{A}}$ and $\varepsilon_{\text{M}}$ are the response amplitude and response mean errors, respectively. A more involved description of the DORA task, files, and data are provided in an open access \href{https://github.com/maneesh51/Benchmark-Tasks/}{GitHub repository}~\cite{YadavStender24}.

\subsection{Passive One-Legged Hopper}
\begin{figure}[btp]
    \centering
    \includegraphics[width=\textwidth]{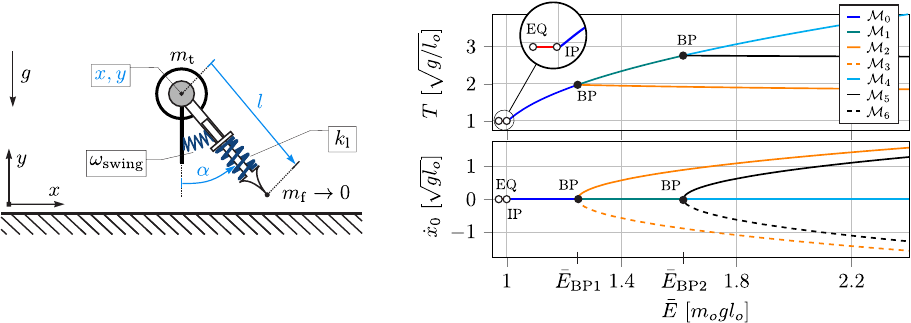}
    \caption{Depicted are the passive one-legged hopper and a bifurcation diagram of its periodic trajectories. Since the system conserves energy, the bifurcation parameter is the energy level~$\bar{E}$. The projections of the periodic motions presented are the period time~$T$ and horizontal velocity~$\dot{x}_0$ at the apex of the hopper. The illustrations are taken from Figure~3 and 4 in \cite{Raff2022}}
    \label{fig:hopper}
\end{figure}

Similar to the Duffing Oscillator presented in the previous section, this dataset is associated with the challenge to learn the dynamics, approximate the first order return map, and predict the bifurcation behavior of a complex nonlinear system.
It is based on the dynamic model of a one-legged hopper, see Figure~\ref{fig:hopper}. 
In comparison to the Duffing oscillator, it stands for a far more complex class of dynamical systems, which are not only nonlinear, but also hybrid and non-smooth and which cannot be modeled using a single set of ordinary differential equations.
Such dynamics greatly increase the \textbf{complexity} associated with simulation and optimization problems.
They are found, for example, in robotic systems that interact with the environment such as in object manipulation or legged locomotion. 
This interaction introduces non-smooth dynamical effects (rapid changes in a system's velocities), necessitating more sophisticated simulation tools that need to detect when such non-smooth effects happen and treat their effects appropriately.

In periodic tasks, such as legged locomotion, the resulting motion is often regarded on the basis of a recurrence map, which computes how the state is updated during one full period.
As a consequence, the resulting simulations are often computationally expensive.
This holds in particular when gradient information must be computed within the simulation to obtain sensitivities.
This derivative information is of great importance not only for the study of the bifurcation behavior, but also for understanding stability and developing efficient and robust controllers for robotic systems.
Despite the high complexity involved in generating time-series data for non-smooth dynamical systems, the resulting dataset can still be low dimensional, which makes it an interesting candidate for learning surrogate functions.

The one-legged hopper is a good example of such a hybrid dynamical system.
It has only two parameters (swing-leg frequency~$\omega_\text{swing}$, leg stiffness~$k_\text{l}$) and six states~$(x,y,\varphi,\dot{x},\dot{y},\dot{\varphi})$. 
Yet, the system exhibits non-smooth behavior when the foot comes in contact with the ground and generating time-series data is thus quite challenging.
As a consequence of the nonlinear dynamics, the hopper also exhibits a rich variety of periodic motions, similar to the Duffing oscillator in the previous section.
A projection of these periodic motions is depicted in the bifurcation diagram shown in Figure~\ref{fig:hopper}.
The hopper also serves as a prominent template model within the fields of biomechanics and robotics~\cite{Blickhan1989,Geyer2006,Gan2018}.
Data-driven predictions on the hopper's dynamics can take one of three forms, depending on how its periodic time-series data is \textbf{interpreted}:
\begin{enumerate}
    \item The evolution of the time-series data can be used to do a sequence prediction or to train a model for the right-hand side of a discrete dynamical system.
    This is similar to how the data in the DORA dataset is handled.
    \item Each time-series in the dataset corresponds to a periodic motion and thus represents a fixed point of a first-return map. Learning such a map significantly simplifies the system dynamics by reducing a recurrent motion to a single point.
    Again, this can be interpreted as a sequence prediction for parameterized dynamics.
    \item The bifurcation diagram can also be learned directly, requiring a model that takes the bifurcation parameter ($\bar{E}$) as its sole input.
\end{enumerate}
These three approaches to training a machine-learning-based model are listed in order of increasing complexity, as the availability of training and testing data decreases significantly for the latter approaches. The \textbf{evaluation} of the dataset depends on the chosen prediction method. For the first approach, the metrics presented in Equations~\eqref{eq:DORA_MSE1}~and~\eqref{eq:DORA_MSE2} are suitable for data evaluation, while the latter two approaches can be assessed by their ability to generate the bifurcation diagram shown in Figure~\ref{fig:hopper}.
Regarding \textbf{generalizability}, the hopper example belongs to the complex class of non-smooth dynamical systems making suitable machine-learning techniques directly applicable to other datasets from this class.

The \href{https://www.doi.org/10.18419/darus-4237}{dataset} is publicly available in DaRUS, the curated data repository of the University of Stuttgart~\cite{Raff2024}.
The model is described in detail in~\cite{Raff2022}. 
This paper is accompanied by \href{https://github.com/raffmax/ConnectingGaitsinEnergeticallyConservativeLeggedSystems}{code} on GitHub~\cite{Raff2022a}. 
The time-series data of periodic motions is provided in comma-separated text files encoding matrices of varying sizes.
The provided dataset description, as well as MATLAB and Python scripts, include the necessary information to access the time-series data and plot the bifurcation diagram in Figure~\ref{fig:hopper}.
Additionally, the non-smooth periodic motions can be interactively visualized with the provided MATLAB live script, allowing for a detailed examination of the bifurcating trajectories.
This allows for a detailed examination of the bifurcating trajectories. 

\subsection{Motion Data of a Four-Bar Mechanism}

% Motivate the problem (short)
Creating virtual models of mechanical systems is a common task in engineering to gain better insights and to develop suitable control algorithms. 
However, this requires careful and extensive parameter identifications of the systems and the resulting models usually do not perfectly describe the behavior due to unknown dynamics, complex state-dependent forces, or other, intentionally unmodeled effects.
Utilizing recorded data from hardware experiments can help to improve the virtual model and better align it with the real-world system.
This motivates the development of pure data-driven methods or hybrid models that combine data-driven and physics-based approaches such as in \cite{Roder2024}.

\begin{figure}[bt]
\centering
\includegraphics[width=0.4\textwidth]{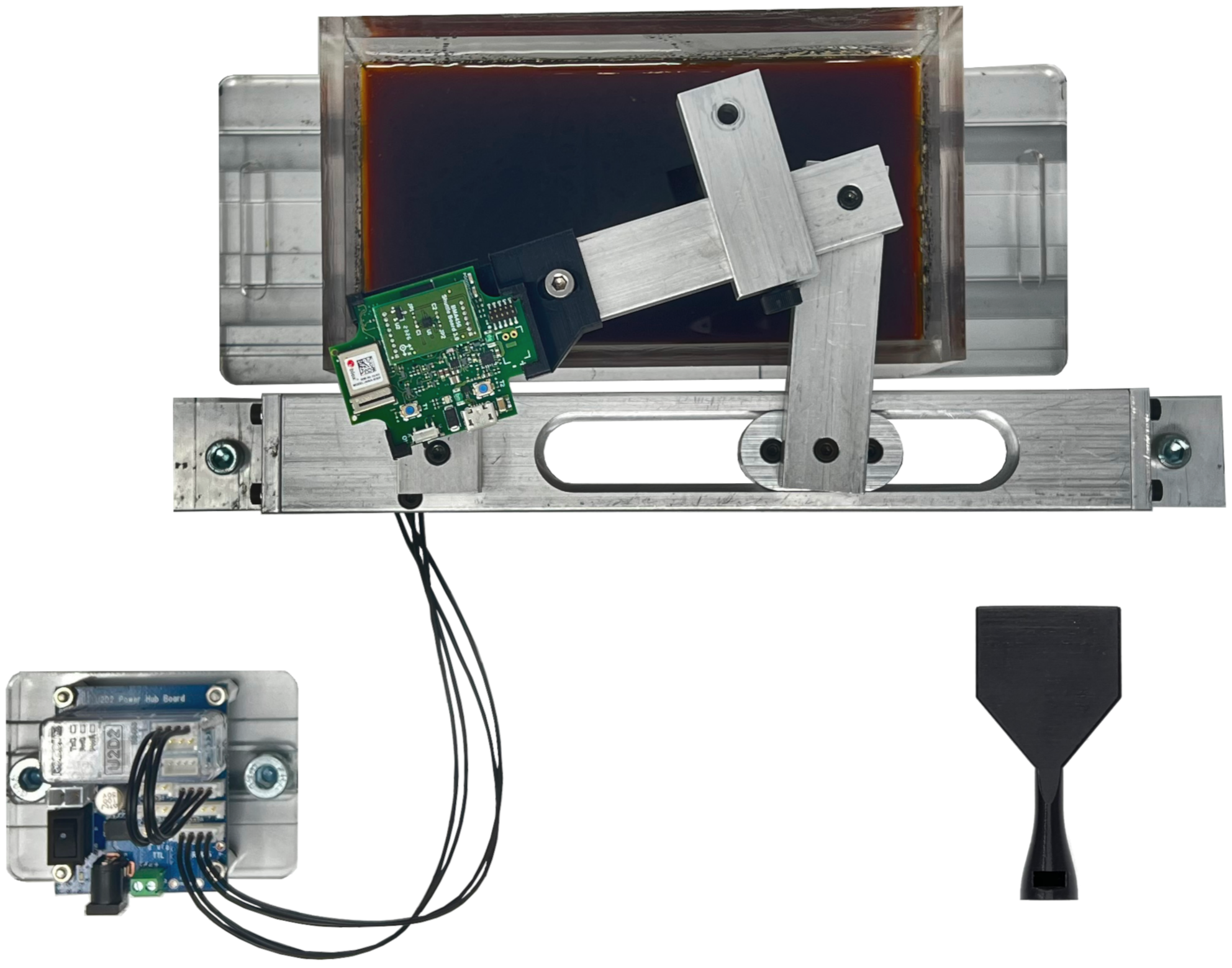}
\caption{Hardware prototype of a four-bar mechanism used to stir a viscous fluid, where the stirrer can be replaced. An exemplary stirrer is shown in the lower right}
\label{fig:fourbar_hardware}
\end{figure}

\begin{figure}[bt]
    \centering
    \begin{subfigure}[b]{0.49\linewidth}
        \centering
        \includegraphics[width=\linewidth]{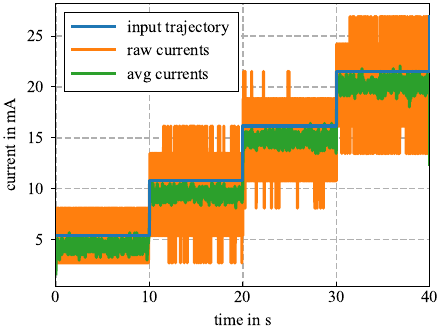}
        \caption{}
        \label{fig:fourbar_data_input}
    \end{subfigure}
    \hfill
    \begin{subfigure}[b]{0.49\linewidth}
        \centering
        \includegraphics[width=\linewidth]{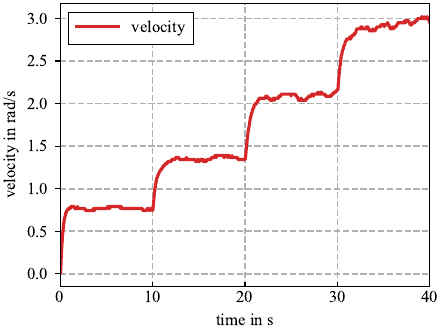}
        \caption{}
        \label{fig:fourbar_data_output}
    \end{subfigure}
    \caption{(a) Prescribed current trajectory and corresponding current measurements. (b) Measured velocity for the applied currents}
    \label{fig:fourbar_data}
\end{figure}

% Explain what the data is and what is special about it (short)
    % Implement a figure here and reference it (check Tutorial notebook)
The presented dataset contains motion and motor-current data from hardware experiments of a four-bar mechanism.
Despite its simplicity, the four-bar linkage is a widely used mechanism, because when carefully designed, it can convert rotational motion into translational motion in a very flexible manner.
The mechanism is driven by a Dynamixel XH430-W350-R motor that allows for current control, i.e., desired current values are supplied, which the on-board electronics strive to reach.
The prescribed current trajectories serve as the input to the system, whereas the measured, actual current, and the resulting angular positions and the angular velocities of the motor shaft are recorded.
Figure~\ref{fig:fourbar_hardware} shows the hardware prototype and Figure~\ref{fig:fourbar_data} shows one of the recorded trajectories. 

By mounting a stirrer onto the mechanism and moving it through a viscous liquid, different damping effects can be introduced into the system.
The complete dataset, a notebook for intuitive handling of the data, and a detailed description of the mechanism are published in a curated data repository of the University of Stuttgart~\cite{darus-4152_2024}. 
They were used by a partner research group to compare two different hybrid-modeling approaches in~\cite{Wohlleben2024}. 
Such an identified model can then be used, e.g., to obtain a model-based controller, furnishing, together with the data-driven modeling process, a control-design assistant. 

% Explain what interesting approaches can be done with it and when considered succesful (is this even the case for me?) (short)
The measurements can be utilized by others to train an end-to-end data-driven model of the system capable of predicting the behavior without any expert knowledge.
Another approach is to build a hybrid model where either an error correction model can be learned or the model parameters are fitted from the data. This can lead to a better understanding of the model and potentially allow insights into the previously unknown dynamics.
Additionally, as with any data from a physical system, the presented data exhibits measurement noise and other disturbances such as coarse measurement resolutions, which are often not dealt with in purely theoretical works on data-driven methods in engineering, but which are a regular occurrence and key limiting factor on real hardware. 
Hence, the dataset can also be used to test different preprocessing strategies for handling noisy and quantized data.

% Comment on how my data deals with the mentioned problems (long)
    % Comment on the Evaluation challenge
    % Conclude with more technical comments on the dataset
For the \textbf{evaluation} of proposed models, a part of the data can be held back as test datasets to assess the prediction accuracy and generalization capabilities.
Various metrics can be defined such as the mean error for a one-step prediction or comparisons of the complete time series data to check for slow drifts over time.
Even though the system is conceptually simple, learning models from hardware data exhibits a high \textbf{problem complexity} due to unknown external influences such as manufacturing inaccuracies, the presence of measurement noise and quantization,  imperfections of the motor unit, and introduced damping stemming from the fluid-structure interaction of the stirrer with a viscous liquid.  
Concerning the \textbf{interpretability}, the dataset itself is easy to understand and is provided in text files which can be quickly loaded into any programming environment. 
Additionally, a python notebook is provided to guide the user through the data and to show how to load and visualize it.
This allows also non-domain experts to develop own data-driven models without any system knowledge on a simple mechanical system commonly used in engineering.
Furthermore, the dataset can be utilized as a benchmark to compare against hybrid models developed by domain experts.
Regarding \textbf{generalizability} and the scarcity of data from hardware experiments, the presented dataset also serves as a starting point for the development and verification of  algorithms applicable to more complex systems.

\subsection{Axial Turbine Dataset}
\label{sec:axial_turbines}
In recent years, the demand for renewable energies has drastically grown to reduce the dependency on fossil fuels. Among all, hydropower itself contributes 47\% of total renewable energies produced in 2023 \cite{Ritchie_Roser_Rosado_2024}. Axial turbines are one of the popular machines for hydropower plants due to their nature to be suitable for low-head conditions. %Hydraulic machines can be designed specifically for one operating condition to be fulfilled. 
Typically, hydraulic machines are designed for an operating range with several operating points simultaneously. In their simplest form, they are designed specifically for one operating condition to be fulfilled. For hydraulic machines, the common objectives are design head ($\Delta h$; energy difference between flow inlet and outlet of the turbine), efficiency ($\eta$), power ($P$), mechanical stresses ($\sigma$) and volume of cavitation ($V_{\textnormal{Cav}}$; the volume of fluid (vapor-filled cavities) below the vapor pressure within the liquid). Hence, multi-objective optimizations are performed to get a finalized shape of the axial turbines \cite{fraas2022sensitivity}. Performing multi-objective optimization of an axial turbine is expensive due to the number of expensive computational fluid dynamics~(CFD) simulations involved.

\begin{figure}[b!] 
    \begin{subfigure}{0.42\textwidth}
        \includegraphics[width=\textwidth]{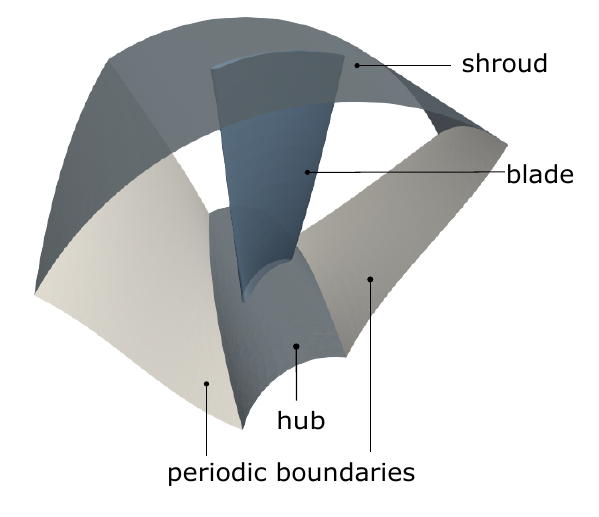}
        \caption{\centering Axial turbine test case}
        \label{fig:tistos}
    \end{subfigure}
    %\hfill
    \begin{subfigure}{0.5\textwidth}
        \includegraphics[width=\textwidth]{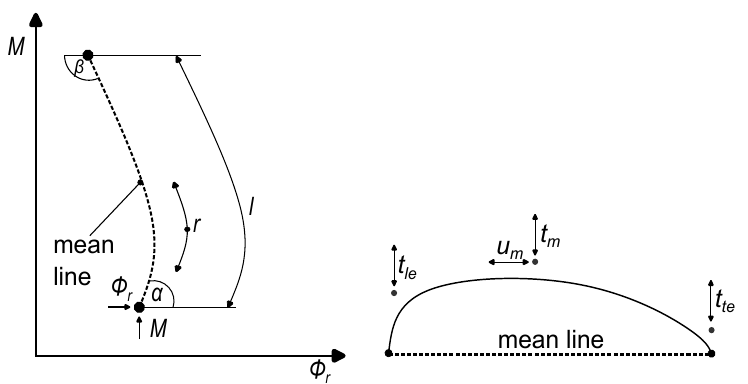}
        \caption{\centering Blade parameterization}
        \label{fig:parameterization}
    \end{subfigure}
    \caption{\centering Section of an axial turbine and the parameterization of the blade \cite{Rentschler2024, Raj2024}}
    \label{fig:axial_turbine_param}
\end{figure}

Figure \ref{fig:axial_turbine_param} visualizes a typical section of an axial machine and the corresponding b-splines to parameterize the blade. The detailed descriptions of the axial turbine and the parameterization are published in \cite{Rentschler2024, Raj2024}. Usually, hydraulic turbines are operated at different operating conditions like part, full, and nominal-loads. Hence during optimization, it is important to take care of operating conditions as well. This means that for each operating point, a separate CFD simulation has to be performed. The overarching optimization target is to obtain high-efficiency values and no cavitating regions for the desired net head and power range. 
%at: do you mean discharge or volume flow? 
%rr: no, I mean to justify the reason behind multiple objectives that what objectives we want to maximize and what we want to minimize

\begin{figure}[bt]
    \begin{subfigure}{0.48\textwidth}
        \includegraphics[width=\textwidth]{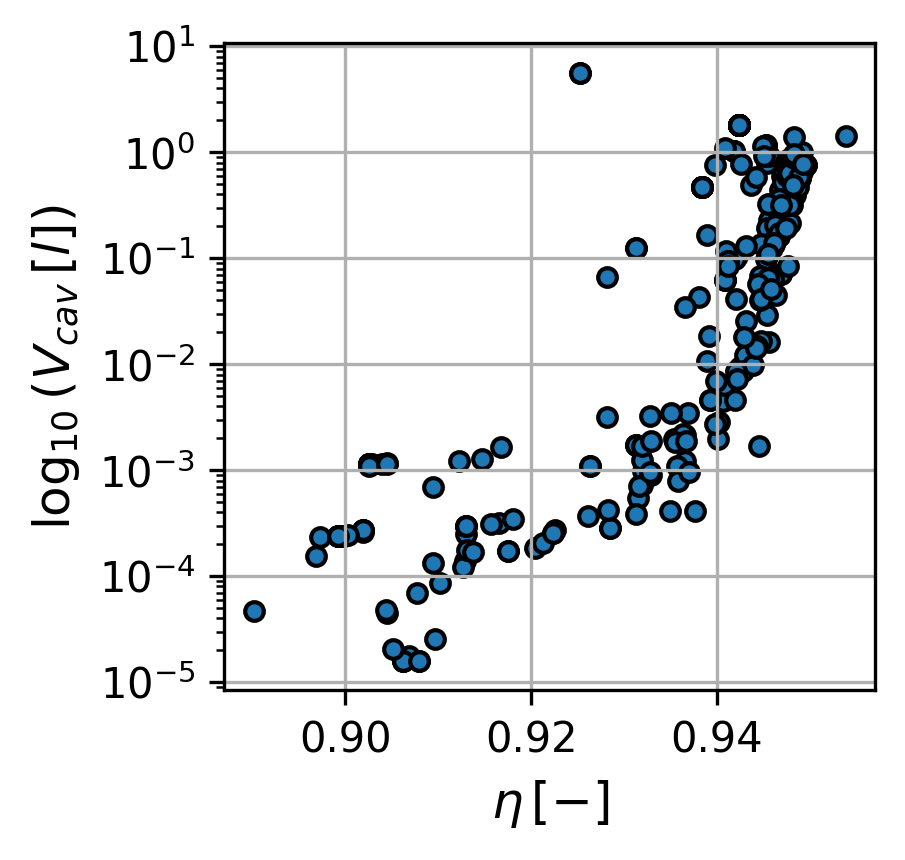}
        \caption{\centering The results present the volume of cavitation versus the efficiency of the machine at nominal load}
        \label{fig:pareto_2d}
    \end{subfigure}
    \hfill
    \begin{subfigure}{0.5\textwidth}
        \includegraphics[width=\textwidth, trim=1.7cm 1cm 0cm 2.7cm , clip]{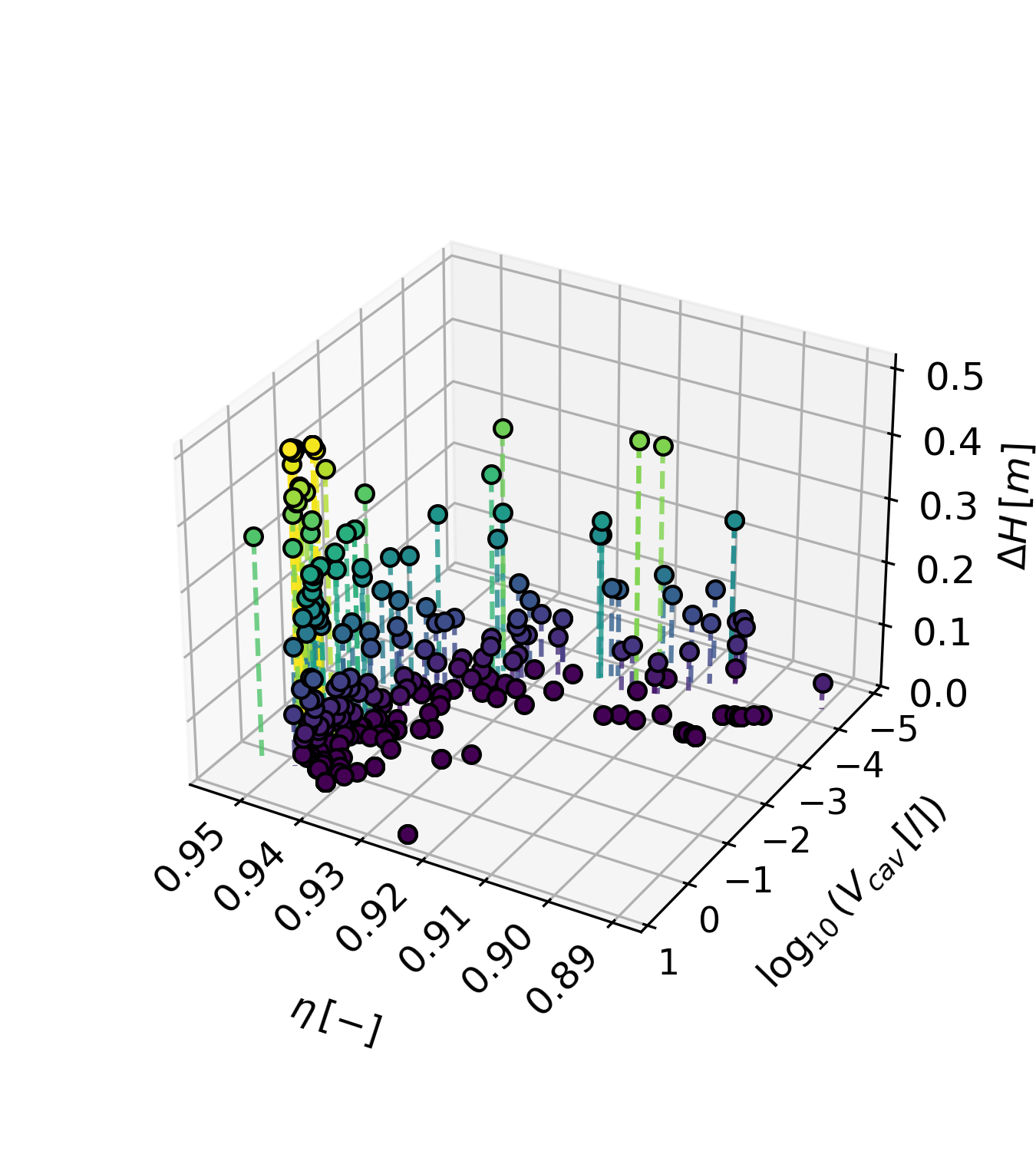}
        \caption{\centering The results present the volume of cavitation versus the efficiency versus design head of the machine at nominal load.}
        \label{fig:pareto_3d}
    \end{subfigure}
    \caption{\centering Pareto front of the dataset after 89 generations}
    \label{fig:Pareto_fronts}
\end{figure}

The dataset presented here is an outcome of an optimization run of an axial turbine shown in Figure~\ref{fig:axial_turbine_param}. The inputs of the dataset consist of 30 geometry parameters, which define the blade at the hub, shroud, and mid-span. The shroud and hub diameter remain constant in the current case. The outputs of the dataset are the corresponding post-processed results of the following CFD simulations. They are power ($P$), design head ($\Delta h$), the efficiency of the machine ($\eta$) and volume of cavitation ($V_{\textnormal{Cav}}$). The presented dataset is publicly accessible at the \href{https://github.com/ihs-ustutt/axial_turbine_database}{Github repository} \cite{axial_turbine_2024} containing geometry and result data in tabular form. Proper descriptions of inputs are given in the open-access article \cite{Raj2024}. 
It is also possible to generate and extend the dataset. The expansion of the dataset for training, validation, and testing will include CFD simulations to perform, which is computationally expensive. The docker container is available with all source code on another \href{https://ihs-ustutt.github.io/dtOO/quickstart.html}{Github repository}, which is based on the design tool (d)esign (t)ool (O)bject-(O)riented~(dtOO) \cite{Tismer_2020, dtOO_2020}. Figure \ref{fig:Pareto_fronts} illustrates the trade-off between different objective functions. 
The presented Pareto front in Figure \ref{fig:pareto_2d} is a two-dimensional projection of the three-dimensional set of non-dominated solutions, displaying the key objectives when the volume of cavitation is minimized at the same time, maximizing the efficiency of the machine.

This comprehensive axial turbine dataset offers valuable resources for both engineers and machine learning practitioners. Engineers can utilize this data to %validate \ac{cfd} models, 
optimize turbine designs, and benchmark performance across various operational conditions. The inclusion of cavitation data is particularly useful for improving turbine longevity and reliability. Machine learning engineers can leverage this dataset to develop predictive models for turbine performance, potentially creating algorithms that can forecast efficiency and cavitation risks based on input parameters. Such models could be invaluable for real-time monitoring and predictive maintenance in hydropower plants. Additionally, the dataset could be used to train deep learning models for automated design optimization, potentially uncovering novel turbine configurations that maximize efficiency while minimizing cavitation.

The dataset of this relatively simple turbine geometry is \textbf{interpretable} with basic fluid mechanics and turbomachinery knowledge. It is available to train machine-learning-based algorithms, specifically mapping from different design parameters to the post-processed results. Regarding \textbf{generalizability}, since hydraulic machines are specifically designed for specific applications, the design or machine type cannot be generalized in the sense of large-scale production. However, this simple case may serve as a training object to optimize more complex turbine types in further work. 
Furthermore, the prediction of the flow field is based on the numerical simulation of the coupled nonlinear, partial differential equations (Navier-Stokes equations), which makes the generalizability in the case of hydraulic turbines impossible. In terms of the \textbf{problem complexity}, the multi-objective problem can be changed to a single-objective optimization. Neglecting other objectives can significantly reduce the complexity of the dataset. However, with the current state-of-the-art turbine design problems, the complexity will substantially increase, primarily because of an increase in the number of geometry parameters. Owing to the flow interaction between several turbine components, the optimization of just one component is not sufficient.  On the contrary, it is essential to design several turbine components simultaneously in order to optimize the entire machine.

\subsection{Lessons Learned}
While translating the perceived challenges and proposed solution approaches of data publishing in the studied area to the seven concrete examples, the authors learned various lessons that can be summarized as follows, focusing on aspects not already discussed at length:
\begin{itemize}
\item Researchers seem to routinely underestimate the time and effort it takes for others to use their data and code.
\item A review process for data publications and associated code is usually necessary to ensure quick accessibility even for other domain experts. The reviewer needs to actually execute the software and load and work with the data to ensure its functionality on an unprepared computer. As this is not usually required for publications of any sort, one should ask fellow scientists to try it and report back how long it took them to make it work. This practice was used here where applicable. 
\item It can indeed be non-trivial to manually or automatically evaluate the quality, e.g., of design results in mechanics and dynamics, different from, e.g., applications of learning methods to natural language and images. 
\item All authors have worked with their own published datasets described here and partly with those of their co-authors to use them within engineering design. Many have experienced that the composition of the data and potential data preprocessing or alternative data representations were usually very fundamental to the performance of the developed approach. 
For many engineering tasks, it is still very unclear which data representation is best suited, unlike for some other fields like, e.g., computer vision where successful ways to represent images for learning have been established. Sometimes, it is even unclear which data of an engineering system to record, to begin with. 
\item Pure publications of datasets in a data repository are nowadays citable with a digital object identifier. However, for key performance indicators used to assess researchers' careers, such as the $h$-index, data publications of this kind and citations of them are often irrelevant. This is in contrast to the undisputed importance of data for data-based methods and the skill and effort it requires to obtain and prepare a useful dataset. This discourages the publication of data for which one is not publishing a paper in parallel that uses this data and neglects the worth of published data in its own right. As papers usually propose functioning solutions, in effect, publication of data for unsolved problems is discouraged. 
\end{itemize}

\section{Conclusions and Outlook}
\label{sec:conclusions}
Data science, machine learning, and artificial intelligence have in recent years expanded from first success stories in computer vision and natural language processing to progress in natural sciences and engineering. 
Regarding the latter, artificial intelligence also starts to have a lasting impact on engineering design, and will transform how humanity conceives new machines, vehicles, robots, etc. 
As this article has outlined, engineering design brings about characteristic perspectives and challenges of which not every aspect and challenge are present for the currently most successful applications of  machine learning. 
In particular, this includes the challenge of \textbf{evaluation}, as for design problems, there usually is no known best or true solution relative to which the quality of results can be gauged. 
Also, computationally evaluating the quality of a proposed design can be very expensive, e.g., by performing expert-driven and computationally highly demanding simulations, whereas data from experiments is usually not yet available in early design stages or very scarce. 
As a consequence, the generation of large datasets becomes prohibitively expensive.  
Even worse, these simulations often benefit from high-performance computing architectures different than the ones most fitting for, e.g., deep learning as engineering simulations typically rely more significantly on general-purpose processing. 
In the future, this may be alleviated, e.g., through simulators that, themselves, are based on machine-learning approaches. 
Nevertheless, when proposing and evaluating radically new designs, there remains the risk that such machine-learned simulators would be evaluated in regions where they have no prior knowledge, which may yield arbitrarily unreliable results, or require currently unseen generalization. 
Still, as the examples provided in this paper show, today's methods and state of the art already allow to publish and use data and data-based approaches in the field of engineering design, giving first answers of how to overcome associated challenges. 
Clearly, researchers in engineering are only at the beginning of transforming engineering design toward a data-driven future. 
And one of the most interesting aspects will be whether this will lead to a future full of auto-designed, highly task-specific machinery that can be extremely efficient and accurate due to designs perfected for the specific task, not limited by the amount of time, money, and knowledge available in the form of human design engineers. 

However, as publication habits change to include ever more data and code, also the question arises whether reviewing processes need to adapt. Does the reviewer of the future need to rate the interpretability of supplied data and ensure that future, competitive approaches can be readily and fairly evaluated? 
Will there be scientifically peer-reviewed publications of data, without accompanying paper and with only the necessary data usage information and processing code, for which citations will count toward key performance indicators for scientific careers? 
Many more questions of this kind will accompany the future of publication habits in mechanics, dynamics, and engineering design. 

\section*{Acknowledgments}
All authors acknowledge the support by the German Research Foundation (Deutsche Forschungsgemeinschaft, DFG) within the Priority Program (DFG SPP) 2353 "Daring More Intelligence: Design Assistants in Mechanics and Dynamics". 
P.\ Eberhard acknowledges support under DFG Grant No.\ 501890093 "Daring More Intelligence – Design Assistants in Mechanics and Dynamics (SPP 2353) - Coordination Proposal", and H.\ Ebel is grateful for the past support from the same grant. 
P.\ Eberhard and B. R\"{o}der acknowledge support under DFG Grant No.\ 501840485. 
J.\ van Delden and T.\ L\"{u}ddecke acknowledge the support by the Deutsche Forschungsgemeinschaft (German Research Foundation), project number 501927736, within the DFG Priority Program 2353. 
They also acknowledge the computing time made available to them on the high-performance computers HLRN-IV at GWDG at the NHR Centers NHR@G\"{o}ttingen. 
These centers are jointly supported by the German Federal Ministry of Education and Research and the German state governments participating in the NHR (www.nhr-verein.de/unsere-partner).
A.\ Borse, R.\ Gulakala, and M.\ Stoffel gratefully acknowledge the financial support provided by Deutsche Forschungsgemeinschaft Priority Program SPP 2353 (DFG Grant No.\ STO 469/16-1). 
M.\ Stender and M.\ Yadav acknowledge the support by the Deutsche Forschungsgemeinschaft under grant 501847579. 
L.\ Schindler and K.\ M.\ de~Payrebrune gratefully acknowledge the funding by the Deutsche Forschungsgemeinschaft (DFG, German Research Foundation) under grant number 501861263. 
M.\ Raff and C.\ D.\ Remy acknowledge the funding by the Deutsche Forschungsgemeinschaft (DFG, German Research Foundation) -- 501862165. M.\ Raff further acknowledges the support through the International Max Planck Research School for Intelligent Systems (IMPRS-IS). 
The contribution of R.\ Raj, T.\ Rentschler, A.\ Tismer, and S.\ Riedelbauch was funded by the German Research Foundation (DFG) (Grant/Award Number: 501932169) and the corresponding simulations were performed on the national supercomputer HPE Apollo Hawk cluster at the High-Performance Computing Center Stuttgart (HLRS) under the grant number parDtOOMeetsCanada/44196. 
The funders had no role in study design, data collection and analysis, decision to publish, or preparation of the manuscript.

\end{document}